\documentstyle[12pt]{article}    
\begin{document}
{\bf Two-loop effective potential in noncommutative scalar field theory\\ }
                                                  
\begin{center}
\vspace{1cm}

                      Wung-Hong Huang\\
                       Department of Physics\\
                       National Cheng Kung University\\
                       Tainan,70101,Taiwan\\

\end{center}
\vspace{2cm}

    The renormalization of effective potential for the noncommutative
scalar field theory is investigated to the two-loop approximation.    It is
seen that the nonplanar diagram does not appear in the one-loop potential. 
  However, nonplanar diagram can become dominant in the two-loop level as
the noncommutativity of geometry is sufficiently small.    The result shows
that the radiative corrections from the  nonplanar diagrams have an
inclination to induce the spontaneously symmetry breaking if it is not
broken in the tree level, and have an inclination to restore the symmetry
breaking if it has been broken in the tree level.

\vspace{3cm}
\begin{flushleft}
 E-mail:  whhwung@mail.ncku.edu.tw\\
 Typeset by Latex
\end{flushleft}

\newpage
\section  {Introduction}

   Noncommutative field theories have been received a great deal of
attention [1-4].     Historically,  it is a hope that the deformed geometry
in the small spacetime would be possible to cure the quantum-field
divergences, especially in the gravity theory.    The work of Filk [5],
however, proved that the noncommutative theory exhibits the same divergence
as the commutative one.  

  The renovation of the interesting in noncommutative field theories is
that it have proved to arise naturally in the string/M theories [6-9].   
Initially,  Connes, Douglas and Schwarz [6]  had shown that the
supersymmetric gauge theory on noncommutative torus is naturally related to
the compactification of Matrix theory [10].   More recently,  it is known
that the dynamics of a D-brane in the presence of  a B-field can, in
certain limits, be described by the noncommutative field theories [9].   

  The quantum aspects of the noncommutative field theories have been
pursued via perturbative analysis over diverse model [11-16].   A distinct
characteristic of the  noncommutative field theories, found by Minwalla,
Raamsdonk and Seiberg [13], is the mixing of ultraviolet (UV) and infrared
(IR) divergences reminiscent of the UV/IR connection of the string theory. 
  For the $\phi ^4$ theory, it has been proved to remain ultraviolet
renormalizable up to two loops [14], although this does not seem to hold to
all order [12].   

    In this paper we will investigate the effective potential for the
scalar field theory in the noncommutative spacetime.   Our interesting is
to see how the noncommutativity of the spacetime will affect the
Coleman-Weinberg mechanism [17].    We will use the path-integration
formulation [18] to evaluate the effective potential to the two-loop
approximation.   In this formulation it is easy to see that there is no
nonplanar diagram in the one-loop potential.   Thus the spontaneous
symmetry breaking is blind to the noncommutativity at this level.  
However, we find that the nonplanar diagram can become dominant in the
two-loop potential, if the noncommutativity of geometry is sufficiently
small.    We also find that the nonplanar diagram has an inclination to
induce the spontaneously symmetry breaking  if it is not broken in the tree
level.   It also  has an inclination to restore the symmetry breaking if it
has been broken in the tree level.

\section  {Formulation  and one-loop diagram}
  The noncommutative geometry we considered is defined by  the coordinates
$x_\mu$ which satisfy the relation  
\
    $$ \left[ x_\mu, x_\nu \right] = i \theta_{\mu\nu} , \eqno {(2.1)}$$
\noindent 
in which $\theta_{\mu\nu}$ is a real, antisymmetric matrix.   The function
defined over the noncommutative spacetime can be expressed as 

$$f(x) = {1\over (2 \pi)^4} \int d^4 k ~ e^{i k_\mu  x^ \mu} \tilde{f}(k) .
\eqno{(2.2)}$$
\
The Moyal product ($\ast$) of two functions is then defined by [11-16]

$$f(x) \ast  g(x) = {1\over (2 \pi)^8} \int d^4 k d^4 p e^{i k_\mu  x^\mu}
e^{i p_\nu x^ \nu} \tilde{f}(k) \tilde{g}(p)~~~~~~~ $$
$$ \hspace{2.5cm} =  {1\over (2 \pi)^8} \int d^4 k d^4 p  e^{i (k_\mu +
p_\mu) x^\mu - {i\over 2} k_\mu \theta^{\mu\nu} p_\nu}  \tilde{f}(k)
\tilde{g}(p)$$
$$= e^{+{i\over 2} \theta^{\mu\nu} {\partial\over \partial y^\mu} 
{\partial\over \partial z^ \nu} } f(y) g(z) |_{y,z\rightarrow x}.   
\eqno{(2.3)} $$

   We consider the $\phi^4$ theory with the action

$$S[\phi] = \int d^4 x ~{\cal L}(\phi) = \int d^4 x ~ [~{1\over 2}
\partial_\mu \phi \ast  \partial^ \mu \phi
+ {1\over 2} m^2 \phi \ast \phi - {\lambda\over 4!}\phi \ast \phi \ast \phi
\ast  \phi ] .  \eqno{(2.4)}$$
\noindent
After expanding the  Lagrangian around a constant field $\phi_0$ the action
can be expressed as 

$$S[\phi] = S[\phi_0] + {1\over 2}\int d^4 x ~ d^4 y ~\tilde{\phi}(x) \ast
\tilde{\phi}(y) ~ \left .{\delta^2 S \over {\delta \phi (x) \delta \phi
(y)}}  \right | _{\phi_0}+ \int d^4 x ~\tilde {\cal L_I} (\tilde \phi ,\phi
_0 ), \eqno{(2.5)}  $$
\
in which $\tilde{\phi} \equiv \phi - \phi_0 $ and $\tilde {\cal L_I}
(\tilde \phi ,\phi _0 )$ can be found from  the Lagrangian Eq.(2.4).   
(Note that as the constant field $\phi_0$ is a stationary point  we have 
the relation $\left .{\delta S \over \delta \phi}  \right | _{\phi_0} = 0.
$)    Then using the propagator defined by  

$$i D^{-1}(\phi_0;k) = \int d^4 k e^{i k x} i D^{-1}(\phi_0;x),
\eqno{(2.6a)}\\$$
$$ i D^{-1}(\phi_0;x,y) = \left .{\delta^2 S \over {\delta \phi (x) \delta
\phi (y)}}  \right | _{\phi_0}, ~~~~~~~~ \eqno{(2.6b)}  $$
\\
the effective potential $V(\phi_0)$ is found to be [18]

$$V(\phi_0) = V_0 (\phi_0) - {1\over 2} i ~\hbar \int {d^4 k \over {(2
\pi)^4}} \ln det i D^{-1}(\phi_0;k) + i ~ \hbar <exp \left ({i \over \hbar
}\int d^4 x ~  \tilde {\cal L_I} (\tilde \phi ,\phi _0 ) \right )>. 
\eqno{(2.7)} $$

The first term in Eq.(2.7) is the classical potential.   The second term is
the one-loop contribution from the second term in Eq.(2.5).  It is known
that, as $\theta_{\mu\nu}$ is an antisymmetric matrix the quadratic part of
the field $\phi(x)$ in the noncommutative spacetime will be identical with
that in the commutative spacetime, after the spacetime integral [11-16].   
Therefore, this elementary property implies that  the Moyal product in the
second term of Eq.(2.5) can be dropped.    Thus we see an interesting
property that the noncommutativity of spacetime dose not affect the
potential in one-loop level.    This property had been found by Campbell
and Kaminsky [15] after investigating the tadpole diagram in the linear
sigma model.

   The third term in Eq.(2.7) is the higher-loop contribution of the
effective potential.   To obtain it we shall evaluate the expection value
of the third term in Eq.(2.5) by the conventional Feynmam rule, with
$D(\phi_0;k)$ as the propagator and keep only the connected single-particle
irreducible graphs [18].      In the next section we will analyze the
two-loop diagram and find that the nonplanar diagram can become dominant if
the noncommutativity of geometry is sufficiently small.      
\section {Two-loop diagram}

    In the two-loop level  there are two planar and nonplanar diagrams.   
Using the Feynman rule, which  includes the propagator

\unitlength 2mm
\begin {picture}(30,5)
\put(15,0){\vector(1,0){5}}
\put(20,0){\line(1,0){5}}
\put(20,1){$p$}
\end {picture}
\hspace{5mm} $1\over p^2 +M^2$

\noindent
and vertices

\begin {picture}(50,12)
\put(15,0){\line(1,1){10}}
\put(15,10){\line(1,-1){10}}
\put(15,0){\vector(1,1){3}}
\put(15,10){\vector(1,-1){3}}
\put(25,0){\vector(-1,1){3}}
\put(25,10){\vector(-1,-1){3}}
\put(17,0){$p_1$}
\put(17,10){$p_4$}
\put(22,0){$p_2$}
\put(22,10){$p_3$}
\put(30,5){$ -  6 \lambda V(p_1,p_2,p_3,p_4)$}
\end {picture}

\begin {picture}(50,12)
\put(15,0){\line(1,1){5}}
\put(20,5){\line(1,-1){5}}
\put(20,5){\line(0,1){5}}
\put(15,0){\vector(1,1){3}}
\put(25,0){\vector(-1,1){3}}
\put(20,10){\vector(0,-1){3}}
\put(17,0){$p_1$}
\put(22,0){$p_2$}
\put(21,8){$p_3$}
\put(30,5){$ -  6 \phi_0 \lambda V(p_1,p_2,p_3)$}
\end {picture}

\noindent
in which  $M^2 \equiv m^2 + {1 \over 2} \lambda \phi _0 ^2$ and  $V(p_i) =
e^{\frac{-i}{2}\sum_{i<j} p_{i\mu} \theta^{\mu\nu} p_{j\nu}}$ , the
contributions of the effective potential from the planar diagram are [18]

\begin {picture}(19,8)
\put(9,2){\circle{4}}
\put(13.2,2){\circle{4}}
\put (17,2){= $I_1^P$}
\end {picture}
$$ = ~{2\over 3}~{{\hbar ^2} \over 24} \lambda \int  {d^4 k \over
{(2\pi)^4}}{1\over k^2 +M^2} \int  {d^4 p \over {(2\pi)^4}}{1\over p^2
+M^2},    \eqno{(3.1)}$$

\begin {picture}(15,8)
\put(11,2){\circle{4}}
\put(9,2){\line (1,0 ){4}}
\put (17,2){= $I_2^P$}
\end {picture}
  $$ \hspace{4cm} = - ~ {1\over 2} ~ {{\hbar ^2} \over 36} \lambda ^2
\phi_0^2\int  {d^4 k d^4 p \over {(2\pi)^8}}{1\over {(k^2 +M^2)( p^2 +M^2)(
(k+p)^2 +M^2)}}.    \eqno{(3.2)}$$

\noindent
The contributions of th effective potential from the nonplanar diagram are
like those in the planar diagram,  while with an extra factor $e^{i k_\mu
\theta^{\mu\nu} p_\nu}$.   They are

$$I_1^N = ~ {1\over 3} ~ {{\hbar ^2} \over 24} \lambda \int  {d^4 k d^4
p\over {(2\pi)^8}}{e^{i k_\mu \theta^{\mu\nu} p_\nu}\over( k^2 +M^2)( p^2
+M^2)},    \eqno{(3.3)}$$

  $$I_2^N = - ~ {1\over 2} ~ {{\hbar ^2} \over 36} \lambda ^2 \phi_0^2\int 
{d^4 k d^4 p \over {(2\pi)^8}}{e^{i k_\mu \theta^{\mu\nu} p_\nu}\over {(k^2
+M^2)( p^2 +M^2)( (k+p)^2 +M^2)}}.    \eqno{(3.4)}$$
\\
Note that the factors ${2\over 3}$ (${1\over 3}$) appearing in Eqs. (3.1)
((3.3)) means that the associated planar (nonplanar) diagram will be with
2/3 (1/3) weight of the commutative graph.   And factors ${1\over 2}$
appearing in Eqs. (3.2) and (3.4) means that the associated planar and
nonplanar diagram will be both with 1/2 weight of the commutative graph.  
The counting  has been detailed by Campbell and Kaminsky [15] in
investigating the linear sigma model.   Let us describe it again for the
completeness.  

  The diagram (3.1) has a single vertex, and so a phase factor
$V(p,k,-k,-p)$.   Of the six possible orderings (modulo cyclic permutation)
of the set $\{p,k,-k,-p\}$, four lead to a trival phase factor, and two
lead to a phase of either $e^{    
i k_\mu \theta^{\mu\nu} p_\nu}$ or $e^{- i k_\mu \theta^{\mu\nu} p_\nu}$
(which are the same under the integral over the loop momenta $k$).  Thus
the planar diagram will with 4/6=2/3 weigth and the nonplanar diagram will
with 2/6=1/3 weigth with respect to the commutative graph.  .  

   The diagram (3.2) has two vertices, and we pick up the phase factor
$V(p,k,-p-k)  V(-k,-p, p+k) $.   Each vertex has two orderings (modulo
cyclic permutation), for four combinations total. Explicity evaluation will
find that two  lead to a trival phase factor, and two lead to a phase $e^{i
k_\mu \theta^{\mu\nu} p_\nu}$.   Thus the planar diagram will with 2/4=1/2
weigth and the nonplanar diagram will with 2/4=1/2 weigth with respect to
the commutative graph.

   The planar-diagram contributions  of Eqs.(3.1) and (3.2) have
ultraviolet divergences and shall be renormalized.  This can be analyzed
following that in [18].  However, due to the phase factor in Eqs.(3.3) and
(3.4) the nonplanar-diagram contributions may be finite.  We will analyze
the nonplanar-diagram contributions in below.

   Let us first analyze Eq.(3.3).    Using the Schwinger parameters,
$\alpha_1$ and $\alpha_2$, we see that 

$$ \int  d^4 k ~ d^4 p {e^{i k_\mu \theta^{\mu\nu} p_\nu}\over( k^2 +M^2)(
p^2 +M^2)} =  \int_0^\infty d\alpha_1 \int_0^\infty d \alpha_2 \int  d^4 k
d^4 p ~ e^{i k_\mu \theta^{\mu\nu} p_\nu} e^{-\alpha_1 ( k^2 +M^2)}
e^{-\alpha_2 ( p^2 +M^2)}\\$$
$$ = \int_0^\infty d\alpha_1 \int_0^\infty d \alpha_2 \int d^4 p ~ e^{- {1
\over {4 \alpha_1}} \tilde p_\mu \tilde  p^\mu} e^{-\alpha_2 ( p^2 +M^2)}
\int  d^4 l ~ e^{-\alpha_1 ( l^2 +M^2)}$$
$$ = \int_0^\infty d\alpha_1 \int_0^\infty d \alpha_2 ~ ({\pi \over
\alpha_1})^2 ~ e^{-(\alpha_1+\alpha_2) M^2}\int d^4 p ~ e^{- {1 \over {4
\alpha_1}} \tilde p_\mu \tilde p^\mu} e^{-\alpha_2 p^2 } ,       
\eqno{(3.5)}$$
\\ 
in which $\tilde p^\mu = \theta^{\mu\nu}p_\nu$ and $l_\mu = k_\mu - {i
\over {2\alpha_1}}\tilde p_\mu $.   
\
Next, we analyze Eq.(3.4).    Using Feynman parameter $w$ and then the
Schwinger parameters, $\alpha_1$ and $\alpha_2$, we see that

$$\int d^4 k ~ d^4 p {e^{i k_\mu \theta^{\mu\nu} p_\nu}\over {(k^2 +M^2)(
p^2 +M^2)( (k+p)^2 +M^2)}}  \hspace{7cm}$$
$$= \int_0^1 dw\int d^4 k ~ d^4 p ~ {1\over { \left[ w (k^2 +M^2)+
(1-w)((k+p)^2 M^2)\right ] ^2 }} ~ {e^{i k_\mu \theta^{\mu\nu} p_\nu}\over
{ p^2 +M^2}}\hspace{3.8cm}  $$
$$=  \int_0^1 dw\int d^4 q ~ d^4 p ~ {1\over { \left[ q^2 + (w-w^2)p^2 +
M^2)\right ] ^2 }} ~ {e^{i q_\mu \theta^{\mu\nu} p_\nu}\over { p^2 +M^2}}. 
\hspace{6cm}$$
$$= \int_0^1 dw \int_0^\infty d\alpha_1 \int_0^\infty d \alpha_2 \int  d^4
q ~ d^4 p ~ \alpha_1 ~ e^{- \alpha_1{ \left[ q^2 + (w-w^2)p^2 + M^2)\right
] }}
 e^{i q_\mu \theta^{\mu\nu} p_\nu}  e^{-\alpha_2 ( p^2 +M^2)}\hspace{2cm}$$
$$ =  \int_0^1 dw \int_0^\infty d\alpha_1 \int_0^\infty d \alpha_2 \int d^4
p ~ ~ \alpha_1~ e^{- {1 \over {4 \alpha_1}} \tilde p_\mu \tilde  p^\mu}
e^{-(\alpha_2+\alpha_1 (w-w^2)) p^2 -\alpha_2M^2} \int  d^4 l ~
e^{-\alpha_1 ( l^2 +M^2)}\hspace{1cm}$$
$$ = \int_0^1 dw \int_0^\infty d\alpha_1 \int_0^\infty d \alpha_2 ~ {\pi^2
\over \alpha_1} ~ e^{-(\alpha_1+\alpha_2) M^2}\int d^4 p ~ e^{- {1 \over {4
\alpha_1}} \tilde p_\mu \tilde p^\mu} e^{-(\alpha_2+\alpha_1 (w-w^2))
p^2},\hspace{2cm}        \eqno{(3.6)}$$
\\ 
in which $q_\mu = k_\mu - (1-w) p_\mu $, $\tilde p^\mu =
\theta^{\mu\nu}p_\nu$ and $l_\mu = q_\mu - {i \over {2\alpha_1}}\tilde
p_\mu $.   To evaluate Eqs.(3.5) and (3.6) furthermore we shall know the
exact form of  $\tilde p_\mu $ which depends on the noncommutativity
parameters $\theta_{\mu\nu}$.  

   Denoting  the noncommutativity parameters as

$$ \theta_{\mu\nu} =\left[ \begin{array}{cccc} 
  0      &  a   &  b   &  c\\
-a & 0          &  d   & e\\
-b  & -d   &     0       & f\\
-c & -e   &  -f  &       0\
\end {array}\right] ,  \eqno{(3.7)}$$
\\
then $\tilde p_\mu \tilde p^\mu \equiv p^\nu \theta_{\mu\nu}
\theta^{\mu\lambda} p_\lambda \equiv  p^\nu U_\nu^\lambda p_\lambda$ where 
$U_\nu^\lambda$ is a symmetric matrix.    Because that any real, symmetric 
matrix can be diagonalized by an orthogonal matrix we can thus change the
orthonormal variables $p_\mu$ to another orthonormal variables $h_\mu$ such
that $ {1 \over 4 \alpha_1} \tilde p_\mu \tilde p^\mu +\alpha_2 p_\mu p^\mu
= \lambda_1 h_1^2 + \lambda_2 h_2^2 + \lambda_3 h_3^2 + \lambda_4 h_4^2$,
where $\lambda_i$ been  the eigenvalues of  the associated matrix.   The
eigenvalues are found to be 

$$\lambda_1=\lambda_2=\alpha_2 + {1 \over 8 \alpha_1}(S +\sqrt {S^2 - 4 D^2
}), ~~~~~ \lambda_3=\lambda_4=\alpha_2 + {1 \over 8 \alpha_1}(S - \sqrt
{S^2 - 4 D^2 }),  \eqno{(3.8)}$$
where
$$S \equiv a^2 +b^2 +c^2+d^2+e^2+f^2 ,   ~~~~~~~~~~ D \equiv   c d- b e + a
f . \eqno{(3.9)}$$

\noindent 
Thus, after the integration of $h_\mu$ Eqs. (3.5) becomes  

$$\int_0^\infty d\alpha_1 \int_0^\infty d \alpha_2 ~ {8 \pi^2 \over S +
\sqrt {S^2 - 4 D^2}+ 8 \alpha_1\alpha_2} ~ {8 \pi^2 \over S - \sqrt {S^2 -
4 D^2}+ 8 \alpha_1\alpha_2} ~ e^{-(\alpha_1+\alpha_2) M^2}. \eqno{(3.10a)}
$$
\\
In a similar way, Eqs. (3.6) becomes  

$$\int_0^1 dw \int_0^\infty d\alpha_1 \alpha_1 \int_0^\infty d \alpha_2 ~
{8 \pi^2 \over S + \sqrt {S^2 - 4 D^2}+ 8 \alpha_1 (
\alpha_2+(w-w^2)\alpha_1)}\hspace{4cm} $$

$$\hspace{3cm} ~ \times {8 \pi^2 \over S - \sqrt {S^2 - 4 D^2}+ 8 \alpha_1(
\alpha_2+(w-w^2)\alpha_1)} ~ e^{-(\alpha_1+\alpha_2) M^2}. \eqno{(3.11a)}$$
\\
\noindent
It is easy to see that the above two relations will become divergent if
$S=D=0$.   This is because that the factors $\frac {1}{8 \alpha_1
\alpha_2}$ and $ \frac {1}{8 \alpha_1( \alpha_2+(w-w^2)\alpha_1)}$ is
divergent at $ \alpha_1=0$ and (or) $\alpha_2=0$.    Therefore when the
noncommutativity parameter $\theta$ is very small we can approximate the
above relations by 

$$(3.10a)\approx \int_0^\infty d\alpha_1 \int_0^\infty d \alpha_2 ~ {16
\pi^4 \over  D^2}  e^{-(\alpha_1+\alpha_2) M^2} = {16 \pi^4 \over  D^2
M^4}. \hspace{3.5cm}\eqno{(3.10b)}$$

$$(3.11a)\approx \int_0^\infty d\alpha_1 \int_0^\infty d \alpha_2 ~
\alpha_1 ~ {16 \pi^4 \over  D^2}  e^{-(\alpha_1+\alpha_2) M^2} = {16 \pi^4
\over  D^2 M^6} . \hspace{3cm}    \eqno{(3.11b)}$$
\\

   Substituting the above results into Eqs.(3.3) and (3.4) we finally find
the nonplanar-diagram contribution of the effective potential

$$ V(\phi_0)^{nonplanar} =  \lambda ~ {\hbar ^2 \over 72} ~ {16 \pi^4 \over
 D^2 M^4}~  ( 1 - {\lambda  \phi_0^2 \over M^2})  \hspace{4cm}
\eqno{(3.12).}$$
\noindent
Thus we see that the nonplanar diagram can become dominant in the two-loop
potential, if the noncommutativity of geometry is sufficiently small.  

   Note that the inverse power of $D^2$ in the Eq.(3.12) is a consequence
of dimensional analysis, coupled with the fact that the diagrams of 
Eqs.(3.1) and (3.2) have degree of divergence 4.   This could guarantee
that the higher loop graphs are not more singular than ${1 \over D^2}$ (up
to logarithmic corrections), whereas they will appear with higher powers of
$\lambda$.  Thus, if $D$ is chosen to be of order $\lambda$, the two loop
diagram will dominate over higher loop diagram (and lower loop diagrms) and
the calculations of this paper become useful in determining the effects of
quantum correction on the symmetry property.  The details of a very similar
argument have presented in reference 13.

   From Eq.(3.13) we have the relation

$$ {\partial V(\phi_0)^{nonplanar}\over \partial (\phi_0^2)} =  \lambda ^2~
{\hbar ^2 \over 9} ~ {\pi^4 \over  D^2 M^8}  ~  ( \lambda  \phi_0^2 -  4
m^2) . \hspace{3cm} \eqno{(3.13)}$$ 
\\
Therefore, if  $ m^2 > 0 $, i.e., the symmetry is not broken in the tree
level, then from Eq.(3.13) we see that the value $\partial
V(\phi_0)^{nonplanar} / \partial (\phi_0^2)$  becomes negative if
$\phi_0^2$ is small.    This means that the nonplanar diagram has an
inclination to induce spontaneously symmetry breaking  if it is not broken
in the tree level.  On the othe hand,  if  $m^2 < 0$, i.e., the symmetry
has been  broken spontaneously broken in the tree level, the value
$\partial V(\phi_0)^{nonplanar} / \partial (\phi_0^2)$  in Eq.(3.13) is
positive definitively.   This means the nonplanar diagram has an
inclination to restore the symmetry breaking if it has been broken in the
tree level. 

   It is difficult to analyze the system with $D=0$ (note that $S$ defined
in Eq.(3.9) is positive), but we belive that the above property will not be
changed.   The problem is remained in futher investigation.
  
\section {Conclusion}

    In this paper we have evaluated  the renormalized effective potential
for the scalar field theory in the noncommutative spacetime to the two-loop
approximation.    In the path-integration formulation we see that there is
no nonplanar diagram in the one-loop potential.   Thus the spontaneous
symmetry breaking is blind to the noncommutativity at this level.  
However, we find that the nonplanar diagram can become dominant in the
two-loop potential, if the noncommutativity of geometry is sufficiently
small.   We also find that the nonplanar diagram has an inclination to
induce spontaneously symmetry breaking  if it is not broken in the tree
level.   It also has an inclination to restore the symmetry breaking if it
has been broken in the tree level.

 Finally, as the U(1) Yang-Mills field will couple to itself in the
noncommutative spacetime [11,16] the new Feynmam diagram shall be involved
in evaluating the effective potential.   It is interesting to see how the
new diagram and the noncommutativity of the spacetime will affect the
Coleman-Weinberg mechanism in the theory including the Yang-Mills field.  
Work on the problem is in progress. 

\newpage
\begin{enumerate}

\item H. S. Snyder, Phys. Rev. {\bf 71} (1947) 38;  {\bf 72} (1947) 68.
\item   Connes, {\it Noncommutative Geometry} Academic. Press, New York,
1994);
Connes A.{\it Gravity coupled with matter and the foundation of non
commutative geometry},
Comm. in Math. Phys. {\bf 182} 155-177 (1996) , hep-th/9603053 . 
\item Landi G. {\it An introduction to noncommutative spaces and their
geometries},
Lecture Notes in Physics, Springer-Verlag , hep-th/97801078 ;
V\'arilly J. {\it An introduction to noncommutative geometry}, Summer
School "Noncommutative geometry and applications", Lisbon, September 1997 ,
physics/97090045 
\item Sch\"ucker T. {\it Geometries and forces}, Summer School
"Noncommutative geometry and applications", Lisbon September 1997 ,
hep-th/9712095 .
\item T. Filk, Phys. Lett. B 376 (1996) 53; see also  Chris P.
Korthals-Altes and A. Gonzalez-Arroyo, Phys. Lett. 131B (1983) 396.
\item  A. Connes, M. R. Douglas and A. Schwarz,
  ``Noncommutative Geometry and Matrix Theory: Compactification on
  Tori'', hep-th/9711162, JHEP 9802:003 (1998).
\item  B.~Morariu and B.~Zumino, ``Super Yang-Mills on the
  Noncommutative Torus,'' hep-th/9807198; 
 C.~Hofman and E.~Verlinde,
``U-duality of Born-Infeld on the Noncommutative Two-Torus,''
hep-th/9810116, JHEP {\bf 9812}, 010 (1998)
\item T.~Krajewski and R.~Wulkenhaar,
``Perturbative Quantum Gauge Fields on the Noncommutative Torus,''
hep-th/9903187;
 M.~M.~Sheikh-Jabbari,
``Renormalizability of the Supersymmetric Yang-Mills Theories on the
Noncommutative Torus'',  hep-th/9903107, 
JHEP {\bf 9906}, 015 (1999).
\item  N.~Seiberg and E.~Witten,
``String Theory and Noncommutative Geometry,'' hep-th/9908142, 
JHEP {\bf 9909}, 032 (1999).
\item N. A. Obers and B. Pioline, Phys. Rep. {\bf 318} (1999) 113. 
\item C.~P.~Martin and D.~Sanchez-Ruiz,
``The One-loop UV Divergent Structure of U(1) Yang-Mills Theory on
Noncommutative $R^4$,'' hep-th/9903077, 
Phys.\ Rev.\ Lett.\  {\bf 83}, 476 (1999)
\item I.~Chepelev and R.~Roiban,
``Renormalization of Quantum field Theories on Noncommutative $R^d$.
I: Scalars,'' hep-th/9911098. 
\item S.~Minwalla, M.~Van Raamsdonk and N.~Seiberg,
``Noncommutative Perturbative Dynamics,''
hep-th/9912072, JHEP {\bf 0002}, 020 (2000).
\item I.~Y.~Aref'eva, D.~M.~Belov and A.~S.~Koshelev,
``Two-loop Diagrams in Noncommutative $\phi^4_4$ Theory'', hep-th/9912075; 
A. Micu, ``Noncommutative $\phi^4$ Theory at Two loop'', hep-th/0008057.
\item B. A. Campbell and K. Kaminsky, Nucl. Phys. B 581 (2000) 240,
hep-th/0003137.
\item A. Matusis, L Ausskind and  N. Toumbas, ``The IR/UV Connection in 
Noncommutative Gauge Theories,''hep-th/0002075 ;
\
M.~Hayakawa,``Perturbative Analysis on Infrared and Ultraviolet Aspects of
Noncommutative QED on $R^4$,'' hep-th/9912167; ``Perturbative Analysis on
Infrared Aspects of Noncommutative QED on $R^4$,'' hep-th/9912094.
\item S. Coleman and E. Weinberg, Phys. Rev. {\bf D 7} (1973) 1888. 
\item R. Jackiw, Phys. Rev. {\bf D 9} (1974) 1686. 

\end{enumerate}
\end{document}